\documentclass[australian,english,prl, singlespace, twocolumn]{revtex4-1}
\usepackage[T1]{fontenc}
\usepackage[latin9]{inputenc}
\usepackage{color}
\usepackage{amsmath}
\usepackage{amssymb}
\usepackage{graphicx}

\makeatletter

\newcommand*\LyXThinSpace{\,\hspace{0pt}}

\@ifundefined{definecolor}
 {\usepackage{color}}{}
\makeatother

\IfFileExists{lmodern.sty}{\usepackage{lmodern}}{}

\makeatother

\usepackage{babel}
\begin{document}
\title{Weak versus deterministic macroscopic realism, and Einstein-Podolsky-Rosen's
elements of reality}
\author{Jesse Fulton, M. Thenabadu, Run-Yan Teh and M. D. Reid}
\affiliation{Centre for Quantum Science and Technology Theory, Swinburne University
of Technology, Melbourne 3122, Australia}
\begin{abstract}
Violation of Leggett-Garg inequalities allows proof of
the incompatibility between quantum mechanics and the combined premises
(called macrorealism) of macroscopic realism (MR) and non-invasive
measurability (NIM). Arguments can be given that the incompatibility
arises because MR fails for systems in a superposition of macroscopically
distinct states $-$or else, that NIM fails.  In this paper, we consider
a strong failure of macrorealism, involving superpositions of coherent
states, where the NIM premise is replaced by the assumption of Bell-locality.
We follow recent work and propose validity of a subset of Einstein-Podolsky-Rosen
(EPR) and Leggett-Garg premises, referred to as \emph{weak macroscopic
realism} (wMR).  In finding consistency with wMR, we identify that
the Leggett-Garg inequalities are violated because of failure of \emph{both}
MR and NIM, but also that \emph{both} are valid in a weaker (less
restrictive) sense. Weak MR is distinguished from \emph{deterministic
macroscopic realism} (dMR) by recognizing that a measurement involves
a reversible unitary interaction that establishes the measurement
setting. Weak MR posits a predetermined value for the measurement
outcome, for the system defined at the time \emph{after} the interaction,
when the measurement setting is experimentally specified. An extended
definition of wMR considers the ``element of reality'' defined
by EPR  for a system A, where one can predict with certainty the
outcome of a measurement on A, by measurement on a system B. Weak
MR posits that the element of reality exists once the unitary interaction
determining the setting at B has occurred. We show compatibility
of systems violating Leggett-Garg inequalities with wMR, but point
out that dMR has been shown falsifiable. We compare wMR with models
put forward elsewhere, and give an argument for wMR, on the basis
that wMR resolves the contradiction pointed out by Leggett and Garg
between failure of macrorealism and assumptions intrinsic to quantum
measurement theory.
\end{abstract}
\maketitle

\section{Introduction}

The interpretation of the quantum superposition of two macroscopically
distinguishable states has been a topic of interest for decades \citep{s-cat-1,cats-coherent-state,frowis,yurke-stoler-1,vlastakis-cats,omran-cats,cat-bell-wang,cats-1,cats-3}.
Schr\"{o}dinger considered a superposition $|\psi_{M}\rangle=\frac{1}{\sqrt{2}}(|a\rangle+|d\rangle)$
where $|a\rangle$ and $|d\rangle$ are macroscopically distinct quantum
states, giving outcomes $a$ and $d$ for a measurement $\hat{M}$
 \citep{s-cat-1}. The outcomes are associated with macroscopically
distinct physical properties, analogous to a cat alive or dead. Schr\"{o}dinger
explained how the standard interpretation given to a quantum superposition
 introduces a paradox when applied to the macroscopic system. The
system is interpreted as being in neither state $|a\rangle$ or $|d\rangle$
prior to measurement $\hat{M}$, suggesting it is somehow simultaneously
in both states $-$ which  would be ``ridiculous'' \citep{s-cat-1}.

Leggett and Garg proposed concrete tests of macroscopic realism versus
quantum mechanics \citep{legggarg-1}. They introduced \emph{macroscopic
realism} (MR) as the premise that ``a system with two macroscopically
distinct states available to it will at all times \emph{be} in one
or other of those states''. Specifically, they assumed the system
to be described by a \emph{hidden variable} $\lambda_{M}$, which
takes the value $+1$ or $-1$ depending on which of the two states
the system is in. The variable is hidden, because quantum mechanics
does not give such an interpretation for the state $|\psi_{M}\rangle$.
In the work of Leggett and Garg, the variable $\lambda_{M}$ specifies
the outcome ($a$ or $d$) for the measurement $\hat{M},$ but not
the outcomes for other more microscopic measurements. Hence, \emph{macroscopic
realism} (MR) posits that the system is in a state with a predetermined
value $\lambda_{M}$ for the measurement $\hat{M}$, but does not
imply the stronger assumption that the system is in one or other of
any quantum state, prior to measurement $\hat{M}$. In order to
test MR, Leggett and Garg introduced the additional assumption of
macroscopic noninvasive measurability (NIM). This assumption however
is generally challenging to justify \citep{weak-lg-measurement,mitchell-budroni,NSTmunro-1,emeryreview,uola-vit-bud,knee-ideal-non-invasive}.
The combined assumptions of MR and NIM are referred to as macrorealism.
Assuming macrorealism, Leggett and Garg derived inequalities, which
are predicted by quantum mechanics to be violated for certain dynamically
evolving systems involving macroscopic superposition states.

There have been many predictions and demonstrations of violation of
the Leggett-Garg inequalities \citep{weak-lg-measurement,mitchell-budroni,NSTmunro-1,emeryreview,uola-vit-bud,jordan_kickedqndlg2-1-2,weak-hybrid-1-1,higgens,wlliams-jordan,asadian,goggin,white-npj,pan,halliwell,dressel,knee-ideal-non-invasive,manushan-cat-lg,robens-1,lauralg-1}.
However, many of these tests rely on microscopic realisations, which
therefore involve the stronger assumption of microscopic realism and
cannot test macroscopic realism. The argument can be put forward that
the violation of the Leggett-Garg inequalities is due to failure of
realism at a microscopic (not macroscopic) level. Macroscopic tests
exist \citep{NSTmunro-1,mitchell-budroni,manushan-cat-lg,lauralg-1,weak-lg-measurement},
but these are susceptible to the criticism that the local measurements
used are in fact invasive \citep{Maroney}, or else involve auxiliary
assumptions e.g. that the ``states'' the system is in, according
to MR, can be prepared in the laboratory, and are therefore describable
as quantum states \citep{NSTmunro-1}.

Motivated by the need to rigorously test macroscopic realism, we examine
in this paper a recently proposed test of macrorealism involving superpositions
of coherent states, namely, entangled cat states \citep{manushan-bell-cat-lg,macro-bell-lg}.
Here, a measurement of the sign $\hat{S}$ of a quadrature phase amplitude
$\hat{X}$ distinguishes between two coherent states $|\alpha\rangle$
and $|-\alpha\rangle$ where $\alpha\rightarrow\infty$, and macroscopic
realism implies the outcome of $\hat{S}$ to be predetermined as either
positive or negative. In this proposal, the question of there being
an invasive measurement is partly resolved, because the measurement
of $\hat{S}$ is made by a spatially separated system $B$, so that
the NIM premise is justified by the Bell's assumption of locality
\citep{bell-chsh,Bell-2,brunner-review,wise-review-bell,bell-reviews,weak-hybrid-1-1}.
The proposal may also be regarded as a macroscopic Bell test, in which
a Bell inequality involving the hidden variables $\lambda_{M}$ is
violated. Such tests violate Bell inequalities for macroscopically
coarse-grained measurements, where it is not necessary to fully resolve
the amplitude $\hat{X}$. While other macroscopic Bell-nonlocality
tests have been put forward \citep{pdd-macro-bell,mdr-bill-prl,mdr-prl-macro,howell-macro,nav-macro,tura,bryan-macro,watts-macro,banaszek-cats,cat-bell-1,wodkiewicz,svetlichny,collins,wildfeuer,macro-cv-bell?,thearle-cv,ketterer,gilchrist,gilchrist-cv-prl,meso-bell-higher-spin-cat-states-bell-1},
the distinction here is that macroscopically distinct states can be
readily identified that allow application of Leggett-Garg's definition
of macroscopic realism (MR). Similar tests involving coarse-grained
measurements have been proposed \citep{jeong-macro-coarse-thermal-cat,jeong-kim-coarse-bell,cv-bell-macro-ali,macro-bell-lg,macro-bell-huang}.

In this paper, our motivation is to examine whether and how it is
possible to obtain consistency with MR, despite that the Leggett-Garg-Bell
inequalities are violated for a macroscopic proposal with a rigorous
justification of non-invasive measurability. In the proposal, it is
clear that the dynamics $U_{\theta}$ associated with the choice of
measurement setting $\theta$ in the Bell experiment plays a key role
in understanding how it is possible to find consistency with MR. The
work of Thenabadu and Reid \citep{manushan-bell-cat-lg} has pointed
out that MR can hold for Bell violations, if defined appropriately
to take into account this dynamics.

Two definitions of MR exist. The definitions depend on whether the
term ``measurement'' includes the unitary interaction $U_{\theta}$
that determines the measurement setting or not. \emph{Deterministic
macroscopic realism} (dMR) posits that the value for the outcome of
the measurement $\hat{M}\equiv\hat{S}$ is specified for the system
prior to the entire measurement process, including that determining
the measurement setting. Here, we consider that the macroscopically
distinct states giving a definite outcome for $\hat{S}_{\theta}$
can be identified for the system prior to the entire measurement process,
including $U_{\theta}$. The analogy is in classical mechanics, where
states with a definite $x$ and $p$ are defined for the system at
any point of time, prior to measurement of either. It has been shown
that dMR is falsifiable, according to quantum predictions \citep{macro-bell-lg,manushan-bell-cat-lg}.

On the other hand, \emph{weak macroscopic realism} (wMR) is a set
of weaker assumptions that are not negated by the macroscopic Bell
inequalities \citep{manushan-bell-cat-lg}. The premise of wMR posits
that \citep{manushan-bell-cat-lg}:

(1) The predetermined value $\lambda_{M}$ describes the system as
it exists after the unitary dynamics $U_{\theta}$, when the measurement
setting is established in the experiment. This means the predetermination
is given for the system prepared with respect to the measurement basis
only. It is assumed also that:

(2) The value $\lambda_{M}$ is not changed retrocausally by any future
measurement or interaction, and is not changed by any spacelike-separated
interactions $U_{\phi}$ or events that might occur at a spatially
separated site.

Recently, Fulton et al \citep{ghz-cat} have extended the definition
of wMR to the situation of the Einstein, Podolsky and Rosen (EPR)
paradox \citep{epr-1}, where one can predict with certainty the result
of a measurement at $A$ by making a measurement at a space-like separated
system $B$. Here, wMR posits that:

(3) The value for the outcome $\hat{S}_{\theta}$ at $A$ is specified
by an ``element of reality'' given by the hidden variable $\lambda_{\theta}^{(A)}$,
once the unitary dynamics $U_{\phi}^{(B)}$ determining the measurement
setting at $B$ has taken place.

Fulton et al point out that wMR can be generalised to apply to the
standard microscopic Bell systems, in which case the premise is referred
to as weak local realism (wLR) \citep{ghz-cat}. This premise is not
negated by Greenberger-Horne-Zeilinger (GHZ) experiments \citep{ghz},
and is not inconsistent with the EPR-Bohm paradox \citep{Bohm-1}.
The EPR paradoxes and GHZ contradictions arise from the strong version
of local realism, where the predetermination is prior to the unitary
interactions, $U_{\theta}^{(A)}$ and $U_{\phi}^{(B)}$ \citep{ghz-cat}.
Joseph et al have applied the wMR and wLR premises to examine realism
in Wigner's friend paradoxes \citep{wigner-friend-macro}.

In this paper, we demonstrate compatibility of the macroscopic Leggett-Garg-Bell
test proposed in \citep{manushan-bell-cat-lg}, with the extended
wMR premises. In Sections II-V, we review the arguments put forward
in \citep{manushan-bell-cat-lg}, and then focus on the wMR premise
(3) involving the assumption of an ``element of reality'', showing
consistency with that premise. Four tests of wMR are proposed, all
of which show consistency between wMR and quantum predictions. In
Section VI, we explain how the wMR premise can be implemented in quantum
measurement theory, where a meter is coupled to a system to allow
a readout of a final value. It was raised by Leggett and Garg that
a system violating macrorealism may not be a good measurement device
\citep{legggarg-1}. Extending arguments put forward in \citep{manushan-bell-cat-lg},
we show how the premise of wMR resolves this paradox of quantum measurement.
We also explain in Section VII how the premise is connected to models
of MR put forward by Maroney and Timpson \citep{Maroney,maroney-timpson}.

\section{Cat states and weak macroscopic realism}

We begin by considering the Schrödinger cat state \citep{yurke-stoler-1,cats-coherent-state,collapse-revival-bec,kirchmair-1}
\begin{equation}
|\psi_{M}\rangle=\frac{1}{\sqrt{2}}\bigl(|\alpha\rangle+i|-\alpha\rangle\bigl)\label{eq:cat-1}
\end{equation}
of a single-mode field $A$. Here $|\pm\alpha\rangle$ are coherent
states with $\alpha$ large and real. These states becomes macroscopically
distinguishable in phase space for large $\alpha$, in analogy with
the ``alive and dead'' states, $|a\rangle$ and $|d\rangle$, of
the ``cat''. Quadrature phase amplitude measurements $\hat{X_{A}}={\color{red}{\color{blue}{\color{black}\frac{1}{\sqrt{2}}}}}(\hat{a}+\hat{a}^{\dagger})$
and \textcolor{black}{$\hat{P_{A}}={\color{red}{\color{blue}{\color{black}\frac{1}{i\sqrt{2}}}}}(\hat{a}-\hat{a}^{\dagger})$
are defined (in a rotating frame) }where $\hat{a}^{\dagger}$,
$\hat{a}$ are mode boson operators ($\hbar=1$) \citep{yurke-stoler-1}.
The states $|\pm\alpha\rangle$ can be distinguished by the measurement
$\hat{M}$ given as $\hat{S}^{(A)}$, which has a value $+1$ if
the outcome of $\hat{X}_{A}$ is positive, and $-1$ otherwise. The
outcomes $+1$ and $-1$ are analogous to the spin outcomes in a Bell
experiment. We refer to $\hat{S}^{(A)}$ as the pointer measurement,
since the value for the outcome can be amplified as in a homodyne
measurement scheme, to give a readout on a macroscopic meter. Here,
we say that the system in the state $|\psi_{M}\rangle$ has been prepared
in a superposition of pointer eigenstates, where the measurement basis
is for the \emph{pointer measurement} $\hat{S}^{(A)}$, since $\alpha$
real. The coherent states become effective pointer eigenstates for
large $\alpha$, since the outcome for $\hat{S}^{(A)}$ is given as
$1$ or $-1$ for $|\alpha\rangle$ and $|-\alpha\rangle$ respectively.
The state $|\psi_{M}\rangle$ is prepared for the pointer measurement
$\hat{S}^{(A)}$ (or $\hat{X}_{A}$), the measurement basis being
eigenstates of $\hat{S}^{(A)}$ (or $\hat{X_{A}}$). 
\begin{figure}[t]
\includegraphics[width=0.5\columnwidth]{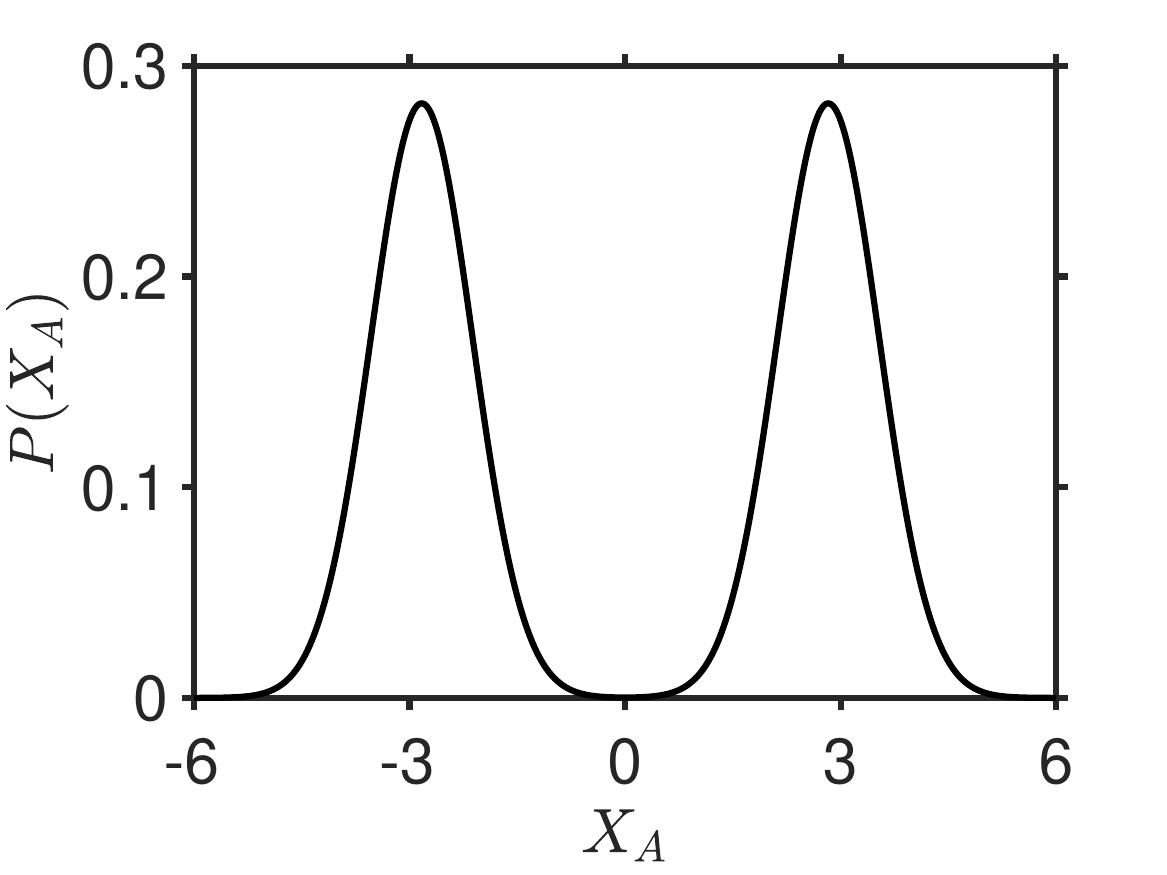}\includegraphics[width=0.5\columnwidth]{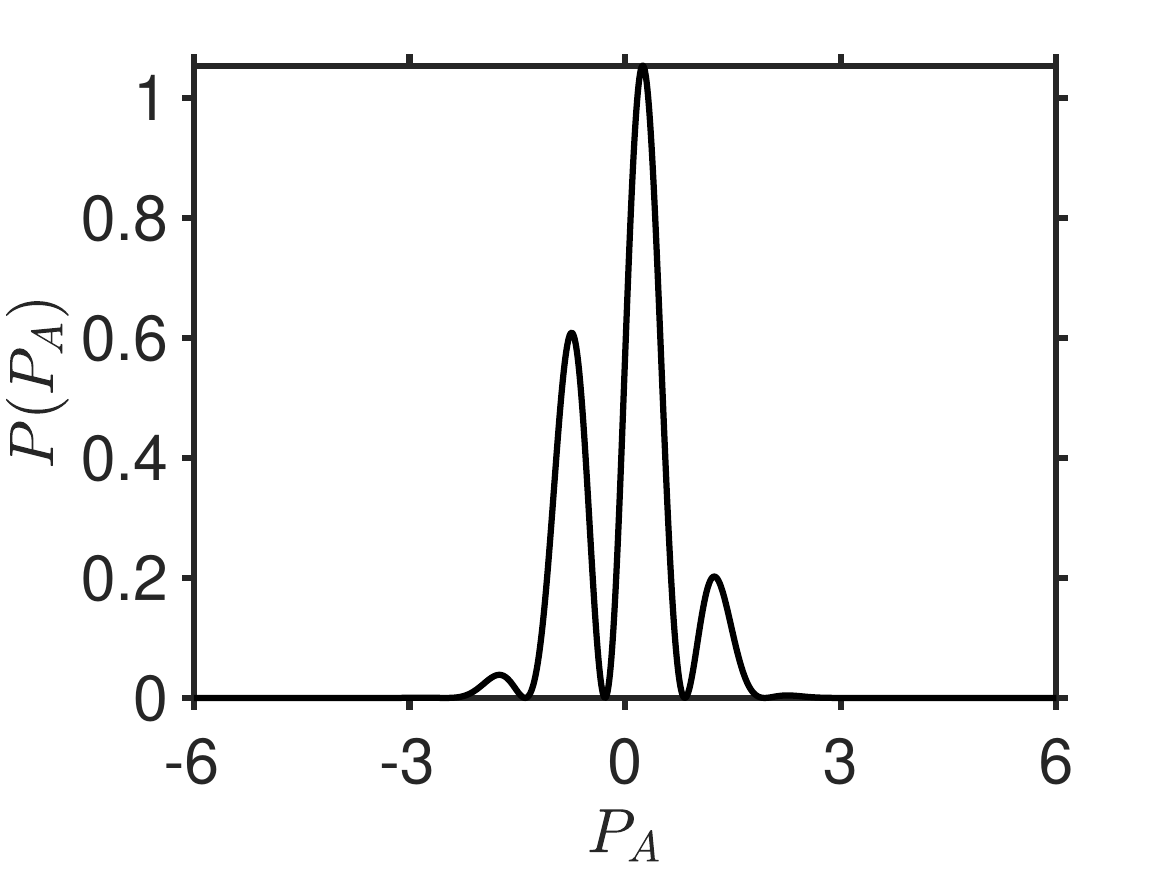}

\caption{The probability distributions $P(X_{A})$ and $P(P_{A})$ for the
cat state (1) with $\alpha=2$.\textcolor{red}{{} }\textcolor{blue}{}\textcolor{red}{\label{fig:probability-distributions}}}
\end{figure}

Before examining whether wMR can be compatible with quantum mechanics,
we present an argument \emph{against} wMR \citep{manushan-bell-cat-lg}.
This concerns the apparent incompleteness of quantum mechanics. Weak
macroscopic realism postulates the system to have a definite ``spin''
outcome, $+1$ \emph{or} $-1$, for $\hat{S}^{(A)}$, which implies
it must be in a state with a sufficiently localised outcome $X_{A}$,
for $\hat{X}_{A}$. If the state is to be a \emph{quantum} state,
we arrive at a constraint on the outcomes $P_{A}$ of measurement
$\hat{P}_{A}$. The distribution for $X_{A}$ gives two Gaussian
hills each with variance $1/2$ (Figure \ref{fig:probability-distributions}).
Supposing the system to be in a classical mixture of two states, one
for each Gaussian, in accordance with wMR, then for each we specify
$\Delta^{2}X_{A}=1/2$. If the two states are quantum states, then
the uncertainty relation $\Delta X_{A}\Delta P_{A}\geq1/2$ for each
implies that the overall variance in $P_{A}$ satisfies $\Delta^{2}P_{A}\geq1/2$
\citep{manushan-bell-cat-lg}. The observation of $\Delta^{2}P_{A}<1/2$
leads to an EPR-type paradox, where, since the state consistent with
wMR cannot also be consistent with the uncertainty principle, one
argues either failure of wMR or else an incompleteness of quantum
mechanics. For the cat state (\ref{eq:cat-1}),\textcolor{red}{{} }a
fringe distribution is observed for $\hat{P}_{A}$ (Figure \ref{fig:probability-distributions}),
and
\begin{equation}
\Delta^{2}P_{A}=\frac{1}{2}-2\alpha^{2}e^{-4\alpha^{2}}{\color{red}}\label{eq:var-p-cat-1}
\end{equation}
The paradox is obtained for all $\alpha$, albeit by a vanishingly
small amount for larger $\alpha$ \citep{macro-coherence-paradox,eric_marg,josa-laura,manushan-bell-cat-lg}.
Similar paradoxes include \citep{irrealism-fringes,bil-angelo-epr}
but are not so directly based on wMR. The apparent inconsistency between
the macroscopic quantum state and the completeness of quantum mechanics
was raised by Schrödinger in his famous essay \citep{s-cat-1}.

While the original EPR paradox revealed inconsistency between local
realism and the completeness of quantum mechanics \citep{epr-1},
Bell later proved local realism could be negated \citep{Bell-2},
giving a resolution of the paradox. The above argument however
is based on weak macroscopic realism (wMR), which motivates the question
of whether wMR can also be negated.

\section{A strong test of macrorealism using entangled cat states}

To examine this question, we first follow \citep{manushan-cat-lg}
to demonstrate how Leggett and Garg's macrorealism \citep{legggarg-1}
can be violated for the cat state (\ref{eq:cat-1}). Macrorealism
is defined as the combined assumptions of wMR and noninvasive measurability
(NIM) $-$ that one may determine the value of $\lambda_{M}$ without
a subsequent macroscopic disturbance to the future dynamics \citep{legggarg-1}.

At time $t_{1}=0$, we consider that system $A$ is prepared in $|\alpha\rangle$.
The system then evolves according to the nonlinear Hamiltonian $H_{NL}=\Omega\hat{n}^{4}$
where $\Omega$ is a constant and $\hat{n}=\hat{a}^{\dagger}\hat{a}$.
After a time $t_{2}=\pi/4\Omega$, it can be shown that the system
is in the state  \citep{manushan-cat-lg,manushan-bell-cat-lg}
\begin{equation}
U_{\pi/8}|\alpha\rangle=e^{-i\pi/8}\bigl(\cos\pi/8\ |\alpha\rangle+i\sin\pi/8\ |-\alpha\rangle\bigl)\label{eq:state2}
\end{equation}
where we write $U_{\pi/8}=U_{A}(t_{2})=e^{-iH_{NL}^{(A)}t_{2}/\hbar}$.
After further evolution, at time $t_{3}=\pi/2\Omega,$ the system
is in the cat state 
\begin{equation}
U_{\pi/4}|\alpha\rangle=\frac{e^{-i\pi/4}}{\sqrt{2}}\bigl(|\alpha\rangle+i|-\alpha\rangle\bigl)\label{eq:second-cat}
\end{equation}
where $U_{\pi/4}=U_{A}(t_{3})$. At each time $t_{i}$, we define
$S_{i}$ to be the outcome of the measurement $\hat{S}^{A)}$. Assuming
the system satisfies wMR, the value for $S_{i}$ is determined by
a macroscopic hidden variable $\lambda_{M}$, which we denote by $\lambda_{i}$,
with values $+1$ and $-1$. Algebra reveals that $\langle\lambda_{1}\lambda_{2}\rangle-\langle\lambda_{1}\lambda_{3}\rangle+\langle\lambda_{2}\lambda_{3}\rangle\leq1$
\citep{legggarg-1,jordan_kickedqndlg2-1-2}. The two-time correlations
are given as $\langle S_{i}S_{j}\rangle=\langle\lambda_{i}\lambda_{j}\rangle$.
The assumption NIM implies these could be measured, since an ideal
measurement of $S_{i}$ at times $t_{i}$ determines the value of
$\lambda_{i}$ without subsequent disturbance to the system. Macrorealism
therefore implies the Leggett-Garg inequality \citep{jordan_kickedqndlg2-1-2,legggarg-1}
\begin{equation}
{\color{black}{\color{black}\langle S_{1}S_{2}\rangle+\langle S_{2}S_{3}\rangle-\langle S_{1}S_{3}\rangle}\leq1}\label{eq:lg-ineq}
\end{equation}
Quantum mechanics predicts $\langle S_{1}S_{2}\rangle=\cos(\pi/4)$
and $\langle S_{1}S_{3}\rangle=0$, since the outcome for $S_{1}$
is known to be $1$ from preparation. Establishing $\langle S_{2}S_{3}\rangle$
is not so clear, because one may argue that a realistic measurement
at time $t_{2}$ will affect the future dynamics. However, assuming
the system is actually in one of the states $|\alpha\rangle$ or $|-\alpha\rangle$
at $t_{2}$, the system at the later time $t_{3}$ will evolve
to $U_{\pi/4}|\alpha\rangle$ or $U_{\pi/4}|-\alpha\rangle$ \citep{legggarg-1}.
This implies $\langle S_{2}S_{3}\rangle=\cos(\pi/4)$. The inequality
(\ref{eq:lg-ineq}) is violated, the left side being $\sqrt{2}$.
Of course, one sees from the paradox (\ref{eq:var-p-cat-1}) that
the system cannot actually quite be in either state $|\alpha\rangle$
or $|-\alpha\rangle$ at time $t_{2}$, prior to measurement.

\emph{}The failure of macrorealism is demonstrated more convincingly,
if one is able to perform the measurement at the time $t_{2}$ without
direct disturbance. To this end, we note the mapping that leads to
the proposal of a macroscopic version of the Bell experiment \citep{manushan-bell-cat-lg}.
As $\alpha\rightarrow\infty$, $|\alpha\rangle$ and $|-\alpha\rangle$
are orthogonal, and we map the system onto spin-qubits $|\uparrow\rangle$
and $|\downarrow\rangle$, defined as eigenstates of Pauli spin $\hat{\sigma}_{z}^{(A)}$.
The rotations $U_{\pi/8}$, $U_{\pi/4}$ and $U_{3\pi/8}$ (defined
below) become precisely the spin rotations required in the Bell experiments,
realised by  Stern-Gerlach analysers or polarising beam splitters
\citep{Bell-2,bell-reviews,bell-chsh}.
\begin{figure}[t]
\begin{centering}
\par\end{centering}
\begin{centering}
\includegraphics[width=1\columnwidth]{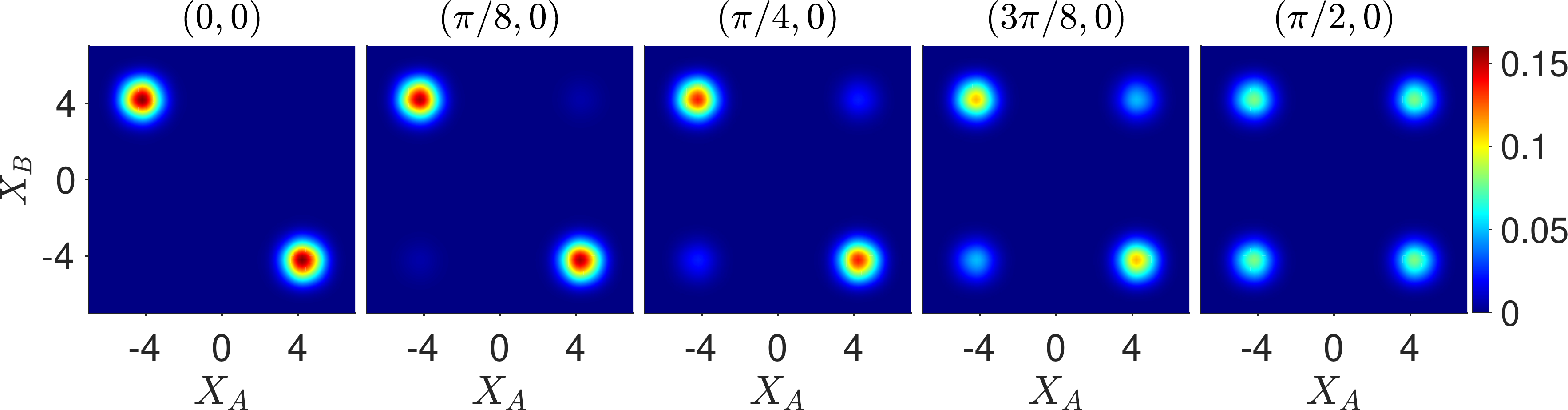}
\par\end{centering}
\smallskip{}

\begin{centering}
\includegraphics[width=1\columnwidth]{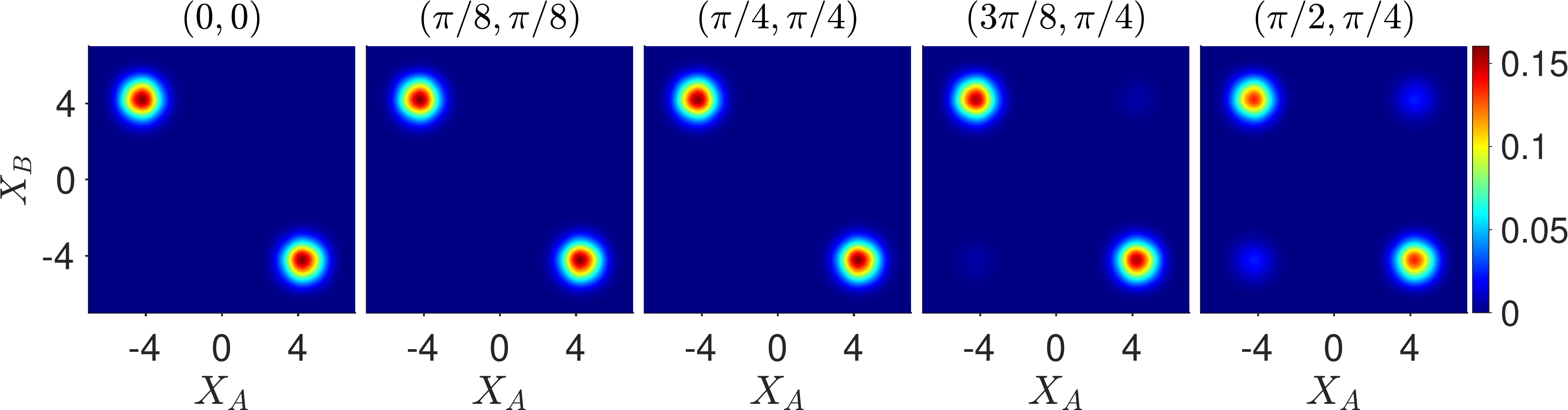}
\par\end{centering}

\caption{Violation of the macroscopic Leggett-Garg-Bell inequality (\ref{eq:ineq})
using cat states: The violation occurs for all $\alpha,$$\beta\rightarrow\infty$.
Contour plots of $P(X_{A},X_{B})$ show the dynamics as the system
prepared in the state $|\psi_{Bell}\rangle_{1}$ at time $t_{1}=0$
evolves through the three measurement sequences of the Leggett-Garg-Bell
test violating (\ref{eq:ineq}). The local systems evolve according
to $H_{NL}^{(A/B)}$ for times $t_{a}$ and $t_{b}$ given by $(t_{a},t_{b})$
in units of $\Omega^{-1}$, for the systems $A$ and $B$. Top: Snapshots
for the measurement of $\langle S_{1}^{(B)}S_{2}^{(A)}\rangle$
and $\langle S_{1}^{(B)}S_{3}^{(A)}\rangle$, where evolution is stopped
at $B$ at time $t_{1}$, so that $t_{b}=0$. This corresponds to
a series of successive unitary rotations occurring at $A$. The interactions
realise $U_{\pi/8}^{A}$, and hence preparation for measurement of
$\hat{S}_{2}^{(A)}$, at time $t_{a}=\pi/4$; and $U_{\pi/4}^{(A)}$,
and hence preparation for measurement $\hat{S}_{3}^{(A)}$, at time
$t_{a}=\pi/2$. Lower: Snapshots for measurement of $\langle S_{2}^{(B)}S_{3}^{(A)}\rangle$,
where evolution is stopped at $B$ at time $t_{2}$, so that $t_{b}=\pi/4$.
This measurement involve two further unitary rotations, and hence
a change of measurement basis at each site $A$ and $B$. Here, $t_{1}=0$,
$t_{2}=\pi/4$ and $t_{3}=\pi/2$. $\alpha=\beta=3$.\textcolor{red}{}\textcolor{red}{}\textcolor{blue}{}\textcolor{red}{\label{fig:contours}}}
\end{figure}

Consider two space-like separated systems $A$ and $B$ prepared at
time $t_{1}=0$ in the state \citep{manushan-bell-cat-lg,cat-bell-wang}
\begin{equation}
|\psi_{Bell}\rangle_{1}=\mathcal{N}\thinspace\bigl(|\alpha\rangle_{A}|-\beta\rangle_{B}-|-\alpha\rangle_{A}|\beta\rangle_{B}\bigl)\label{eq:bell}
\end{equation}
where $|\beta\rangle_{B}$ is a coherent state for system $B$, $\mathcal{N}=\frac{1}{\sqrt{2}}\{1-\exp(-2\left|\alpha\right|^{2}-2\left|\beta\right|^{2})\}^{-1/2}$
and we will take $\alpha=\beta$ with $\alpha$ real. We define the
operators $\hat{X}$, $\hat{P}$, $\hat{S}$, $\hat{n}$, $H_{NL}$,
$U_{\pi/8}$, $U_{\pi/4}$ and hidden variables $\lambda_{i}$ as
above for each system $A$ and $B$, denoting by a superscript $A$
or $B$ in each case. \textcolor{black}{}The systems $A$ and $B$
evolve independently for times $t_{a}$ and $t_{b}$, respectively,
according to \emph{local} interaction Hamiltonians $H_{NL}^{(A)}$
and $H_{NL}^{(B)}$. We define $S_{j}^{(A)}$ ($S_{j}^{(B)}$) as
the outcomes of the measurements $\hat{S}_{j}^{(A)}$ ($\hat{S_{j}}^{(B)}$)
of $\hat{S}^{(A)}$ ($\hat{S}_{j}^{(B)}$) performed after an interaction
time $t_{a}=t_{j}$ ($t_{b}=t_{j}$). If both systems evolve
for a time $t_{2}=\pi/4\Omega$, the system is in the Bell state $|\psi_{Bell}\rangle_{2}=U_{\pi/8}^{(A)}U_{\pi/8}^{(B)}|\psi_{Bell}\rangle$
given by
\begin{eqnarray}
|\psi_{Bell}\rangle_{2} & = & \mathcal{N}e^{-i\pi/4}\bigl(|\alpha\rangle|-\beta\rangle-|-\alpha\rangle|\beta\rangle\bigl)\label{eq:bell2}
\end{eqnarray}
If both systems evolve for a time $t_{3}=\pi/2\Omega$, the system
is in the similar Bell state $|\psi_{Bell}\rangle_{3}=U_{\pi/4}^{(A)}U_{\pi/4}^{(B)}|\psi_{Bell}\rangle$.
The premise wMR assigns to $A$ and $B$ after interaction
times $t_{a}=t_{i}$ and $t_{b}=t_{j}$ ($i,j=1,2$ or $3$)  the
macroscopic hidden variables $\lambda_{i}^{(A)}$ and $\lambda_{j}^{(B)}$.
These take values $+1$ or $-1$ that determine the outcomes for
$\hat{S}_{i}^{(A)}$ and $\hat{S}_{j}^{(B)}$.

The $S_{i}^{(A)}$ can be measured, by taking $t_{b}=t_{i}$ and inferring
the value from a measurement of $S_{i}^{(B)}$ at $B$. The anti-correlation
evident in the Bell states implies $S_{i}^{(A)}=-S_{i}^{(B)}$, and
$\lambda_{i}^{(A)}=-\lambda_{i}^{(B)}$. It is argued that this measurement
is noninvasive to the system $A$, based on the assumption of macroscopic
Bell locality (ML). ML asserts that for spacelike-separated events
at $A$ and $B$, the events at $B$ cannot change the value of the
hidden variable $\lambda_{M}^{(A)}$ at $A$, and vice versa. This
is assumed for \emph{all} events over the time interval $t_{1}$ to
$t_{3}$, implying no macroscopic changes to the outcomes at $A$
at any time $t_{i}$ due to measurement at $B$ \citep{bell-trajectories-1}.
Assuming macrorealism, the inequality (\ref{eq:lg-ineq}) becomes
the Bell inequality \citep{Bell-2}
\begin{equation}
-\langle S_{1}^{(B)}S_{2}^{(A)}\rangle-\langle S_{2}^{(B)}S_{3}^{(A)}\rangle+\langle S_{1}^{(B)}S_{3}^{(A)}\rangle\leq1\label{eq:ineq}
\end{equation}
The predictions based on the measurements $\hat{X}_{A}$ and
$\hat{X}_{B}$ are calculated by evaluating $P(X_{A},X_{B})$ (Figure
\ref{fig:contours}). Where $\alpha,\beta>1$, the predictions for
$-\langle S_{i}^{(B)}S_{j}^{(A)}\rangle$ are indistinguishable from
those of $\langle S_{i}S_{j}\rangle$ given above for the predictions
of the Leggett-Garg inequality (\ref{eq:lg-ineq}). Violation of (\ref{eq:ineq})
is predicted, as for (\ref{eq:lg-ineq}), the left side being $\sqrt{2}$.
\emph{}The violations are valid for arbitrarily large $\alpha$,
$\beta$ and falsify the \emph{combined} assumptions of wMR and ML.
Hence, one cannot conclude violation of wMR directly.

\section{Falsifying deterministic macroscopic realism}

However, one may falsify \emph{deterministic macroscopic realism}
(dMR). The inequality given by (\ref{eq:ineq}) is seen to be a macroscopic
version of Bell's original inequality \citep{Bell-2,bell-chsh}, applied
to \emph{macroscopic} spin observables $\hat{S}_{j}^{(A)}$ and $\hat{S}_{j}^{(B)}$.
The choice between two times of evolution for each system $A$ and
$B$ (e.g. $t_{2}$ and $t_{3}$ for $A$, and $t_{1}$ and $t_{2}$
for $B$) corresponds to a choice between two measurement settings
(e.g. $\theta_{2}$ and $\theta_{3}$ for $A$, and $\phi_{1}$ and
$\phi_{2}$ for $B$). This choice of unitary rotation $t_{j}$
maps in the microscopic Bell experiment to a choice of analyzer setting
$\theta_{j}$. The Bell inequality (\ref{eq:ineq})  can be derived
assuming \emph{deterministic macroscopic realism }(dMR) \citep{manushan-bell-cat-lg}:
each system $A$ and $B$ is \emph{simultaneously} predetermined to
be in one or other of two macroscopically distinct states, prior to
the choice of measurement setting, so that \emph{two} macroscopic
hidden variables (e.g $\lambda_{2}^{(A)}$ and $\lambda_{3}^{(A)}$
for $A$, and $\lambda_{1}^{(B)}$ and $\lambda_{2}^{(B)}$ for $B$)
are ascribed to each system at the time $t_{1}$. This assumption
naturally incorporates ML, since it is specified that $\lambda_{j}$
cannot change over the course of the unitary dynamics associated with
the adjustment of measurement setting, at either site. Violation of
(\ref{eq:ineq}) falsifies dMR.

We may also consider evolution of (\ref{eq:bell}) for the time $t_{4}=3\pi/4\Omega$,
in which case the evolved state is 
\begin{equation}
U_{3\pi/8}^{(A)}|\alpha\rangle=e^{i3\pi/8}\bigl(\cos3\pi/8|\alpha\rangle+i\sin3\pi/8|-\alpha\rangle\bigl)\label{eq:bell-three}
\end{equation}
This allows evaluation of the familiar Clauser-Horne-Shimony-Holt
Bell inequality \citep{bell-chsh,bell-reviews}
\begin{equation}
|\langle S_{1}^{(B)}S_{2}^{(A)}\rangle+\langle S_{2}^{(A)}S_{3}^{(B)}\rangle+\langle S_{3}^{(B)}S_{4}^{(A)}\rangle-\langle S_{1}^{(B)}S_{4}^{(A)}\rangle|\leq2\label{eq:bell-chsh-1}
\end{equation}
which can also be derived from dMR. A violation is predicted, the
left side being $2\sqrt{2}$. The system prepared in the Bell state
(\ref{eq:bell}) thus evolves after a time $t_{4}$ to the Bell state
$|\psi_{Bell}\rangle_{4}=U_{3\pi/8}^{(A)}U_{3\pi/8}^{(B)}|\psi_{Bell}\rangle_{1}$.
This leads to predictions $\langle S_{3}^{(B)}S_{4}^{(A)}\rangle=-\cos\pi/4$
and $\langle S_{1}^{(B)}S_{4}^{(A)}\rangle=-\cos3\pi/4$, and a violation
of (\ref{eq:bell-chsh-1}), the left side being $2\sqrt{2}$. Eq.
(\ref{eq:bell-chsh-1}) can be viewed as the Leggett-Garg inequality
\begin{equation}
\langle S_{1}^{(A)}S_{2}^{(A)}\rangle+\langle S_{2}^{(A)}S_{3}^{(A)}\rangle+\langle S_{3}^{(A)}S_{4}^{(A)}\rangle-\langle S_{1}^{(A)}S_{4}^{(A)}\rangle\leq2\label{eq:lg-second}
\end{equation}
derived in \citep{legggarg-1} . Similar to (\ref{eq:ineq}), to obtain
(\ref{eq:bell-chsh-1}) we justify the NIM premise using ML, and put
$S_{i}^{(B)}=-S_{i}^{(A)}$ for times $t_{1}$ and $t_{3}$, based
on the anti-correlation of the spins for the Bell states. Alternatively,
Eq. (\ref{eq:bell-chsh-1}) is seen to be a macroscopic Bell inequality,
where one measures the correlation $E(\theta_{i},\phi_{j})=\langle S_{i}^{(A)}S_{j}^{(B)}\rangle$.

\section{Finding consistency of the Leggett-Garg-Bell violations with weak
macroscopic realism}

We now ask whether one can reconcile the violations of macrorealism
and deterministic macroscopic realism (dMR) with the validity of weak
macroscopic realism (wMR). For the Leggett-Garg-Bell tests, it is
clear that the systems are indeed prepared in the pointer superposition
$|\psi_{Bell}\rangle_{i}$ at the time $t_{i}$. \emph{If} weak macroscopic
realism (wMR) holds, then the value of the $S_{i}^{(A/B)}$ \emph{are}
given by $\lambda_{i}^{(A/B)}$ at the time $t_{i}$, in which case
the violations must arise because the non-invasive measurability (NIM),
as justified by locality (ML), premise breaks down. Consistency with
wMR is possible, because the unitary dynamics
\begin{equation}
U=e^{-iH_{NL}t/\hbar}\label{eq:unitary}
\end{equation}
(which in the Bell test gives the choice of measurement setting) has
a \emph{finite time duration}. This is evident in the Figure \ref{fig:contours},
which plots the dynamics given by $U$. The dynamics transforms the
Bell state $|\psi_{Bell}\rangle_{1}$ prepared in the pointer basis
of $\hat{\sigma}_{z}$ at time $t_{1}=0$, into a different Bell state
$|\psi_{Bell}\rangle_{2}$ at time $t_{2}=\pi/4$ (prepared with respect
to different basis ), and then into a different state $|\psi_{Bell}\rangle_{3}$
at $t_{3}=\pi/2$ (prepared in the basis of $\hat{\sigma}_{y}$).
The system given by the state $|\psi_{Bell}\rangle_{1}$ is not viewed
to be \emph{simultaneously} in all three pointer superpositions. One
is therefore able to postulate wMR, without requiring to assume dMR,
which fails, by violation of (\ref{eq:lg-ineq}), (\ref{eq:ineq})
and (\ref{eq:bell-chsh-1}).

Examination of the dynamics associated with the measurement settings
for the Leggett-Garg-Bell tests reveals features consistent with
wMR. We first summarise two tests that were explained in \citep{manushan-bell-cat-lg,delayed-choice-cats}.

\subsection{Test 1: Unitary rotations are required at both sites to display the
macroscopic nonlocality}

Any theory for which wMR is valid predicts that it is the dynamics
involving a unitary rotation at \emph{both} sites that yields the
violation of the inequalities (\ref{eq:lg-ineq}) and (\ref{eq:ineq}).
A similar analysis holds for the violation of (\ref{eq:bell-chsh-1}).
 To show this, we examine the top sequence of Figure \ref{fig:contours}.
The system is prepared in the pointer-measurement basis at time $t_{1}$.
A unitary rotation giving a change of measurement basis then takes
place at $A$ but not $B$. According to wMR, the system at time $t_{2}$
given by snapshot ($\pi/4,0$) can be specified by two variables $\lambda_{2}^{(A)}$
and $\lambda_{1}^{(B)}$ that simultaneously determine the outcomes
$S_{2}^{(A)}$ and $S_{1}^{(B)}$ of the measurements $\hat{S}^{(A)}$
and $\hat{S}^{(B)}$, if performed at time $t_{2}$. Importantly,
$\lambda_{1}^{(B)}=-\lambda_{1}^{(A)}$ also determines the outcome
$S_{1}^{(A)}$ of measurement $\hat{S}^{(A)}$ at time $t_{1}$ (given
by $(0,0)$). The outcome for $S_{1}^{(A)}$ can be determined at
time $t_{2}$ without further unitary rotation, because this is given
by the pointer measurement at $B$ at time $t_{2}$. Hence, wMR (3)
applies. We obtain a description for measurements made at times $t_{1}$
and $t_{2}$ that is consistent with macrorealism.
\begin{figure}[t]
\begin{centering}
\par\end{centering}
\begin{centering}
\includegraphics[width=1\columnwidth]{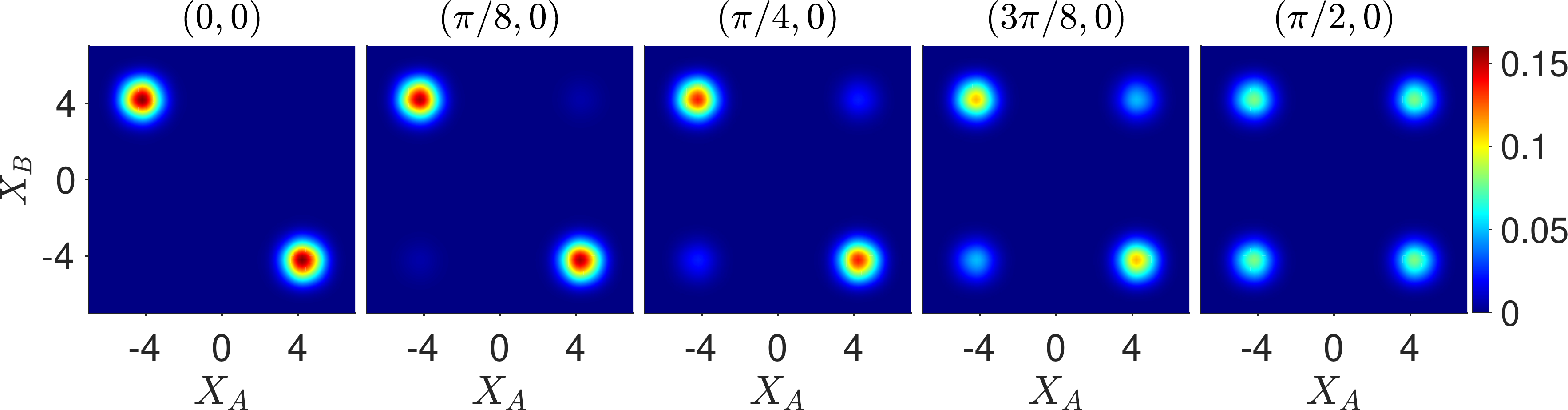}
\par\end{centering}
\smallskip{}

\begin{centering}
\includegraphics[width=1\columnwidth]{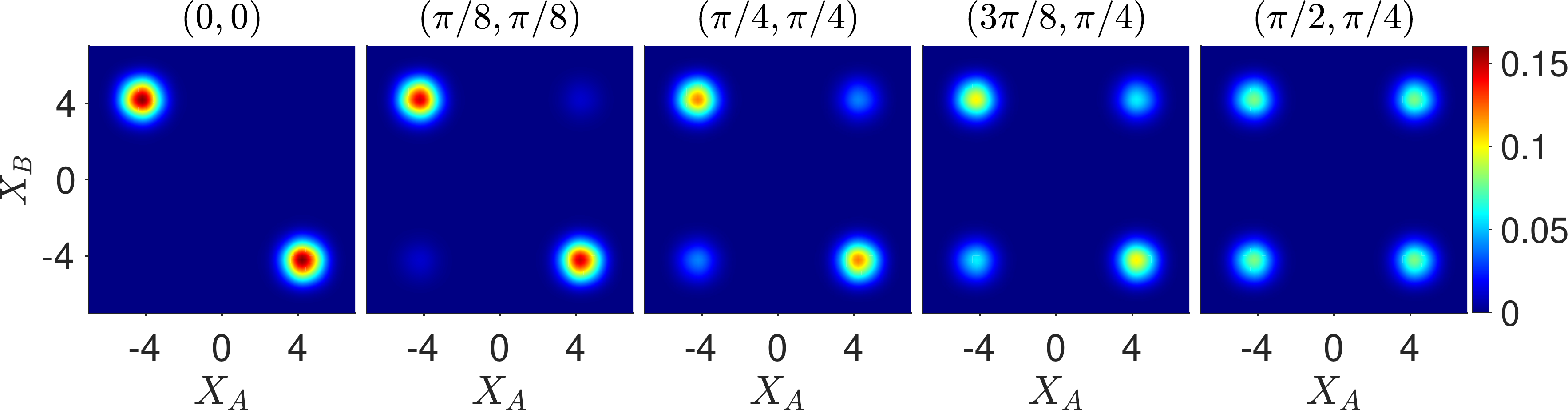}
\par\end{centering}
\caption{Testing weak macroscopic realism by comparing the dynamics with that
of a mixed state: The contours show the sequences associated with
the measurements needed to test the Leggett-Garg-Bell inequality (\ref{eq:ineq})
as described for Figure \ref{fig:contours}, except here the initial
state is taken to be the non-entangled mixed state $\rho_{mix}^{(AB)}$
where no violation is possible. The top sequence is for measurement
of \textcolor{red}{}\textcolor{red}{}\textcolor{blue}{}\textcolor{red}{}$\langle S_{1}^{(B)}S_{2}^{(A)}\rangle$
and $\langle S_{1}^{(B)}S_{3}^{(A)}\rangle$, where a unitary rotation
creating a change of measurement basis takes place at system $A$
only. The lower sequence shows the measurement dynamics for $\langle S_{2}^{(B)}S_{3}^{(A)}\rangle$
involving unitary rotations and hence a change of measurement basis
for both systems, $A$ and $B$. Comparing with Figure \ref{fig:contours},
we see that there is no visual difference between the plots at the
initial time $t=0$. The top sequence involving a rotation at one
site only remains visually indistinguishable from that of the entangled
state shown in Figure \ref{fig:contours}. However, we see that the
final state of the lower sequence involving a change of basis at each
site becomes macroscopically different. \label{fig:mixtures}}
\end{figure}

By contrast, the lower sequence of Figure \ref{fig:contours} has
rotations giving a change of measurement basis for \emph{both} $A$
and $B$. For the system given by $(\pi/2,\pi/4)$ at time $t_{3}$,
wMR asserts validity of variables $\lambda_{3}^{(A)}$ and $\lambda_{2}^{(B)}$
that determine the outcomes of $\hat{S}_{3}^{(A)}$ and $\hat{S}_{2}^{(B)}$,
but there is no determination of the outcome of $\hat{S}_{1}^{(A)}$.
It is the fact that the \emph{three} variables $\lambda_{1}^{(A)}$,
$\lambda_{2}^{(A)}$ and $\lambda_{3}^{(A)}$ cannot be specified
simultaneously for system $A$ at the times $t_{i}$ that allows violation
of the Leggett-Garg inequality. This is not possible, because for
the bipartite system it is only possible to prepare the systems in
pointer bases for \emph{two} measurements simultaneously (one at each
site).

Test 1 provides a way to test wMR. The dynamics for the mixed state
\begin{equation}
\rho_{mix}^{(AB)}=\frac{1}{2}(|\alpha\rangle|-\beta\rangle\langle\alpha|\langle-\beta|+|-\alpha\rangle|\beta\rangle\langle-\alpha|\langle\beta|)\label{eq:mix}
\end{equation}
for which a macrorealistic model holds, can be experimentally compared
with that of $|\psi_{Bell}\rangle_{1}$. Here, $|\alpha\rangle$ and
$|\beta\rangle$are coherent states for systems $A$ and $B$, and
we take $\alpha=\beta$. Weak macroscopic realism predicts that the
dynamics between the two will diverge where there are unitary rotations
at both sites. According to wMR, a violation would not arise where
there are rotations at single sites only. Quantum mechanics predicts
for such an experiment consistency with wMR. Figure \ref{fig:mixtures}
shows the dynamics of the measurements required to test the Leggett-Garg-Bell
inequality for the system prepared initially at time $t_{1}$ in $\rho_{mix}^{(AB)}$.
The top sequence involving a rotation (change of measurement basis)
at one site only is visually unaltered between the cat state $|\psi_{Bell}\rangle_{1}$
(Figure \ref{fig:contours}) and the mixture $\rho_{mix}^{(AB)}$.
The difference between the plots is of order $e^{-\alpha{}^{2}}$
which vanishes for the macroscopic case, where $\alpha=\beta\rightarrow\infty$\citep{supmat}.
By contrast, for the lower sequences where there is a rotation (change
of basis) at both sites, the $P(X_{A},X_{B})$, while indistinguishable
at $t_{1}=0$, become \emph{macroscopically different} at the later
times $t_{2}$ and $t_{3}$. This is seen when comparing the final
plots of the lower sequences: the contour plot for the evolution of
$\rho_{mix}^{(AB)}$ (Figure \ref{fig:mixtures}) is clearly different
to that of $|\psi_{Bell}\rangle_{1}$ (Figure \ref{fig:contours}).

In a model where wMR is valid, it is the dynamics that occurs over
the time intervals of the combination of both unitary rotations involving
a change of measurement basis at each site that results in the violation
of the macroscopic Leggett-Garg-Bell inequalities (\ref{eq:ineq})
and (\ref{eq:bell-chsh-1}). This is consistent with calculations
for violations of the Bell inequalities for microscopic spin Bell
states, where it is well known that the quantum interference arising
from a nonzero angles $\theta$ and $\phi$ is necessary to create
the violation of the inequality (\ref{eq:bell-chsh-1}).

\subsection{Test 2: Delaying the collapse stage of the measurement makes no difference}

The second test of wMR concerns  the timing of the irreversible
``collapse'' stage of measurement, when the system s coupled to
a detector to read out the value of $\hat{X}$. \emph{}We address
how the results are affected by the irreversible ``collapse'' stage
of measurement, when the system $B$ is coupled to a detector. The
unitary evolution $U$, which precedes the collapse stage of the
measurement, prepares the system for the pointer measurement at the
time $t_{i}$, by establishing the measurement setting (i.e. measurement
basis). To calculate the measurable probabilities, we note that in
quantum mechanics, the state for the system is written as a superposition
of pointer eigenstates. 

In a model where wMR is valid, the hidden variable $\lambda_{i}^{(A)}$
is fixed in value ($+1$ or $-1$) at the time $t_{i}$, after $U^{(A)}$.
This gives a record of $\lambda_{i}^{(A)}$ at the time $t_{i}$ $-$
which cannot be changed by the future collapse at $B$, nor by a future
unitary evolution at $A$. 

Quantum mechanics predicts consistency with wMR: It is possible to
delay the collapse stage of the measurement $\hat{S}_{j}^{(B)}$ at
$B$ by any time after the measurement at $A$ i.e. after $t_{3}$,
and it makes no detectable difference to $P(X_{A},X_{B})$, any corrections
being of order $e^{-|\alpha|^{2}}$ \citep{manushan-bell-cat-lg}.As
above, the macroscopic nonclassical effects only arise where there
is unitary rotation at both sites.

\subsection{Test 3: Considering delayed--choice, no-retrocausality implies extra
dimensions}

The premise of wMR specifies a given value $\lambda_{i}$ for the
outcome of the measurement $\hat{S}_{i}$ at the given time $t_{i}$.
This cannot be changed by any future event. One might hence expect
that delayed-choice experiments would falsify wMR.

We note that the joint probabilities $P(X_{A},X_{B})$ depend on the
local interaction times $t_{a}$ and $t_{b}$, not the relative timing.
Hence, one can delay the choice $t_{b}$ to measure $S_{1}^{(A)}$
or $S_{2}^{(A)}$ until \emph{after} the final detection at system
$A$, at time $t_{3}$. This might suggest therefore that the
measurement at $B$ is noninvasive of the dynamics at $A$, and hence
that violation of macrorealism is due to failure of wMR. However,
as with delayed-choice experiments for spin-qubits \citep{wheeler,delayed-choice-kim,delayed-choice-qubit,delayed-choice-rmp,delayed-choice-walborn},
this interpretation can be countered \citep{delayed-choice-causal-model-chaves}:
Analysis reveals unitary evolution $U$ occurs at \emph{both} sites
after time $t_{2}$: the violation of macrorealism can be then explained
by failure of dMR \citep{manushan-bell-cat-lg,delayed-choice-cats}.

On the other hand, the delayed-choice Wheeler-Chaves-Lemos-Pienaar
experiment \citep{delayed-choice-causal-model-chaves,delayed-choice-experiment-chaves-1,huang-delayed-choice-causal-model-compatibility-1}
falsifies all two-dimensional non-retrocausal models for a two-state
system, described by qubits $\{|\uparrow\rangle,|\downarrow\rangle\}$.
Using the mapping (\ref{eq:state2}) onto macroscopic qubits $\{|\alpha\rangle,|-\alpha\rangle\}$
and generalising to rotations $U_{\theta}^{(A)}$ where $\theta$
is a multiple of $\pi/8$, one may falsify all two-dimensional non-retrocausal
models based on the macroscopic qubits \citep{delayed-choice-cats}.
This contradicts wMR $-$ but \emph{only} if we restrict to two-dimensional
models. We avoid conclusions of retrocausality however, by noting
the \emph{extra dimensions} associated with the continuous-variable
phase-space representation of the cat-states, which are measurable.
This is explained in \citep{delayed-choice-cats}.

\subsection{Test 4: EPR's elements of reality are justified after the setting
dynamics}

We now consider the postulate (3) of wMR in the set-up of the EPR
experiment. This extends the earlier work of \citep{manushan-bell-cat-lg}.

Examining the state (\ref{eq:bell}), we see that the Bohm-EPR paradox
for spin applies. At the given time $t_{i}$, the outcome of the measurement
$\hat{S}_{i}^{(A)}$ at $A$ can be predicted with certainty by the
measurement of $\hat{S}_{i}^{(B)}$ on system $B$. We see that $S_{i}^{(A)}=-S_{i}^{(B)}$.
EPR's original premises posit that there exists an \emph{element of
reality} $\lambda_{i}^{(A)}$ for system $A$ at this time \citep{epr-1}.
This value predetermines the outcome of the measurement $S_{i}^{(A)}$
if measured directly at $A$, \emph{regardless of whether the measurement
at $B$ is performed or not}, because the outcome at $A$ can be predicted
in principle by establishing the measurement at $B$ and nothing at
$B$ can influence the system at $A$, according to locality. The
value for the element of reality is $\lambda_{i}^{(A)}=-\lambda_{i}^{(B)}$,
and can be determined by finalizing the measurement at $B$ i.e. making
a readout at $B$.

The original EPR premises can be falsified, because the assumption
of the $\lambda_{i}^{(A)}$ can be applied to (non-commuting) measurements
at the different times, $t_{i}$ and $t_{j}$, as explained in Section
IV. Hence, the EPR premises leads to the premise of dMR which is falsified
by the Bell test violating inequalities (\ref{eq:ineq}) or (\ref{eq:bell-chsh-1})
\citep{Bell-2}.

However, the weaker postulate of wMR (3) is not falsified, because
it refers to the system $B$ at the time $t_{i}$ \emph{after} any
appropriate unitary interaction $U^{(B)}$ has taken place at $B$,
to finalise the measurement setting at $B$ i.e. to prepare the system
for the final pointer measurement $\hat{S}^{(B)}$. This suggests
that the ``elements of reality'' do apply, in certain circumstances.
As explained by Clauser and Shimony \citep{bell-reviews}, the importance
of the dynamics associated with the choice of measurement setting
was commented on by Bohr, in his reply to Einstein, Podolsky and Rosen
\citep{bohr-epr}.

Figure \ref{fig:contours} depicts the dynamics showing consistency
of the quantum predictions with the premise wMR (3). At the time $t_{1}=t_{0}=0$,
the system $B$ has been prepared for the final detection and readout
of spin $\hat{S}_{1}^{(B)}$, which takes place by the measurement
$\hat{X}_{B}$ and a detection. The wMR postulate is that the system
$B$ has the predetermined value $\lambda_{i}^{B}$ for the outcome
of that measurement. This also gives the value for the measurement
$\hat{S}_{1}^{(A)}$at $A$, regardless of any further unitary interactions
$U_{j}^{(A)}$ that might take place at $A$. Indeed, $\lambda_{i}^{(A)}=-\lambda_{i}^{(B)}$.
The Figure \ref{fig:contours} (top) depicts the situation where the
system at $A$ is prepared for the measurement of spin $\hat{S}_{1}^{(A)}$,
without a unitary rotation being required at $A$. It would be possible
to rotate the measurement basis at $A$ to prepare for the measurement
$\hat{S}_{2}^{(A)}$, by evolving the system $A$ according to $U_{\pi/4}^{(A)}$,
but keeping $B$ unchanged. This creates a new state. However, this
does not change the ``element of reality'' $\lambda_{1}^{(A)}=-\lambda_{1}^{(B)}$
for the outcome of measurement $\hat{S}_{1}^{(A)}$. The measurement
$\hat{S}_{1}^{(A)}$ can still be made, by reversing the unitary operation
$U_{\pi/4}^{(A)}$ and performing the final part of the measurement,
$\hat{X}_{A}$. In other words, after a further time, the system evolves
according to the dynamics $(U_{\pi/4}^{(A)})^{-1}$, and the prediction
according to quantum mechanics is that the results of the measurements
at $A$ and $B$ remain anti-correlated. The value $\lambda_{1}^{(B)}$
gives the prediction for $\hat{S}_{1}^{(A)}$ at $A$, regardless
of any local reversible unitary interactions, such as $U_{\pi/4}^{(A)}$
at $A$.

An experimental test of this premise wMR (3) (and of quantum mechanics)
can be performed. The value for the element of reality $\lambda_{1}^{(B)}$
can be obtained by a final measurement readout at the space-like-separated
system $B$. This gives the prediction for $\hat{S}_{1}^{(A)}$. The
value for $\hat{S}_{1}^{(A)}$ can be confirmed correct, both without
and then with the unitary rotation $U_{\pi/4}^{(A)}$ followed by
its reversal.

\section{Weak macroscopic realism, weak local realism and quantum measurement}

The wMR-model can elucidate the nature of the measurement. This can
be put forward as an argument in favour of wMR. A fundamental question
is how to understand the connection between ``realism'' and the
states such as
\begin{equation}
|\psi_{M}\rangle=\frac{1}{\sqrt{2}}(|\uparrow\rangle_{A}|\beta\rangle_{B}-|\downarrow\rangle_{A}|-\beta\rangle_{B})\label{eq:ent-m}
\end{equation}
formed at a time $t_{k}$ after a macroscopic measurement device interacts
with a microsystem $A$ prepared in 
\begin{equation}
|\psi\rangle_{A}=\frac{1}{\sqrt{2}}(|\uparrow\rangle_{A}-|\downarrow\rangle_{A})\label{eq:sup}
\end{equation}
Here, $|\beta\rangle$ and $|-\beta\rangle$ are coherent states.
The readout of $\hat{S}^{(B)}$ gives the measured value of $\hat{\sigma}_{z}^{(A)}$.
A model for an interaction $H$ which evolves (\ref{eq:sup}) into
(\ref{eq:ent-m}) has been presented \citep{model-measurement,laura-measurement-weak-model,blais-meter-model}.
In that model, the meter system is prepared initially in a coherent
state $|\gamma\rangle$. The phase of $\beta$ is determined by the
phase $\gamma$ of the initial coherent state.

A fundamental question arises: At what point in the measurement process
does the value emerge? A deeper question is: How is realism connected
to measurement \citep{bell-against,bell-unspeakable}? 

\subsection{Weak local realism}

It becomes apparent that the wMR premises can also be applied to
spin systems such as $\{|\uparrow\rangle,|\downarrow\rangle\}$, even
though the states are not macroscopically distinct. This is because
the systems violating Bell and Leggett-Garg inequalities for $\{|\uparrow\rangle,|\downarrow\rangle\}$
are a direct mapping of those considered in this paper for $\{|\alpha\rangle,|-\alpha\rangle\}$.
In this case, we refer to the premises of weak macroscopic realism
as simply \emph{weak local realism} (wLR). An explanation has been
given in \citep{ghz-cat}. These premises are weaker (less restrictive)
than those of local realism defined by Bell and are not negated by
violations of Bell inequalities, as we have seen for wMR. We note
in view of Section V.C, this suggests a model in which the spin states
be completed by extra dimensions.

\subsection{Schrödinger's paradox}

Understanding the entangled state (\ref{eq:ent-m}) was the paradox
put forward by Schrödinger in his essay \citep{s-cat-1}. It is
often supposed that the value of the outcome is not determined prior
to measurement, but in the wMR and wLR models, this is overstated.
In these models, the value for the outcome of spin $\hat{\sigma}_{z}^{(A)}$
of $A$ is specified at or by the time $t_{k}$ of the creation of
the entangled system-meter state (\ref{eq:ent-m}), since the \emph{measurement
basis (setting) for the meter is specified by the interaction}, through
the phase of the coherent field. Only direct amplification and detection
of $\hat{X_{B}}$ of the meter is required to complete the measurement.
Hence, according to wMR, there is a predetermined value $\lambda_{M}^{(B)}$
for the macroscopic meter $B$ at this time. According to wMR (3),
this value is an ``element of reality'' for the outcome of $\hat{\sigma}_{z}^{(A)}$at
$A$, since it gives the value if it were to be measured directly.
Hence, in the wMR and wLR models, the value for the outcome of the
measurement can be assigned to the system $A$ at time $t_{k}$, prior
to the final detection and readout, since the measurement setting
for $A$ has been established.

A local unitary interaction at $A$ can be further applied, to change
the measurement setting for the spin measurement at $A$. However,
we have seen that in the wMR-wLR models, this makes no difference
to the outcome $\lambda_{M}^{(B)}$ specified for spin $\hat{\sigma}_{z}^{(A)}$
at $A$, as given by the ``element of reality'' defined at $B$.
The result for spin $\hat{\sigma}_{z}^{(A)}$ is specified by the
meter, and would be verified if the measurement $\hat{\sigma}_{z}^{(A)}$
at $A$ is actually performed. If a local unitary interaction has
changed the measurement setting at $A$, then for the spin $\hat{\sigma}_{z}^{(A)}$
to \emph{actually} be measured, a further unitary interaction giving
a reversal then takes place.

Similarly, if a local unitary interaction at $B$ is implemented,
while keeping the system $A$ unchanged, it does not change the element
of reality for the system $B$ that is implied by the fact the outcome
$\hat{X}_{B}$ can be inferred by the spin measurement at $A$. On
the other hand, if unitary interactions are implemented to change
the measurement settings at both $A$ and $B$, then in the wMR-wLR
models, we can no longer suppose that the value of $\lambda_{M}^{(B)}$
applies to a future measurement.

\subsection{Leggett and Garg's paradox}

In their paper \citep{legggarg-1}, Leggett and Garg consider states
such as (\ref{eq:ent-m}). They explain that the violations of macrorealism
should ``\emph{not be formally in conflict with the arguments so
often given in discussions of the quantum theory of measurement to
the effect that once a microsystem has interacted with a realistic
measuring device, the device  (and, if necessary, the microsystem)
behave as if it were in a definite (and noninvasively measurable)
macroscopic state}''.

They also suggest that a system $|\psi_{M}\rangle$ if violating macrorealism
would not be a suitable measuring device, by continuing: ``\emph{The
macroscopic systems suitable for a macroscopic quantum coherence experiment
are certainly not able to be measuring devices, at least under the
conditions specified. But such a result might cause us to think a
great deal harder about the significance of ``as if''!}''

We extend the analysis of the statements of Leggett and Garg for this
system, given in \citep{manushan-bell-cat-lg}. We examine the first
statement of Leggett and Garg. The premise of wMR does imply a definite
``state'' for the measuring device, given by system $B$ in (\ref{eq:ent-m}),
in the sense that there is a predetermination of the outcome of $\hat{S}^{(B)}$.
This is because the measurement setting $\hat{S}^{(B)}$ is specified
by the phase of the coherent-state amplitude $\beta$.

Assuming wLR, the microsystem $A$ also has a definite value for
the outcome of $\hat{\sigma}_{z}^{(A)}$ $-$ but only when prepared
(after the choice of measurement setting) in a superposition with
respect to the pointer bases of $\hat{\sigma}_{z}^{(A)}$ and $\hat{S}^{(B)}$.
When we write the original state $|\psi_{A}\rangle$, it is not specified
whether or not the measurement basis has been determined experimentally.
However, we see as explained above that once entangled with the meter
as in the state (\ref{eq:ent-m}), there is a definite value for the
outcome of $\hat{\sigma}_{z}^{(A)}$. This is because the measurement
setting for the microsystem is specified.

Hence, there is no direct conflict with the statement ``that the
device behaves \emph{as if} it were in a definite macroscopic state''.
The basis for the spin at $A$ is determined fixed as $\hat{\sigma}_{z}^{(A)}$
because the coherent states that act as the meter (when $\hat{X}_{B}$
is measured) have a definite fixed phase, and no further rotation
$U^{(B)}$ is necessary. The value $\lambda_{M}^{(A)}$ for $\hat{\sigma}_{z}^{(A)}$
is determined by that of $\lambda_{M}^{(B)}$. According to the analysis
of the previous section, there is an element of reality $\lambda_{M}^{(A)}$
for the result of spin $\hat{\sigma}_{z}^{(A)}$ of $A$, for the
system in the entangled meter-system state, and this value is not
changed by any further unitary interactions that may occur at system
$A$.

Now we turn to examine Leggett and Garg's second statement. By mapping
$\{|-\alpha\rangle,|\alpha\rangle\}$ onto $\{|\uparrow\rangle,|\downarrow\rangle\}$
for (\ref{eq:bell}) in Section III, we see that macrorealism is
indeed violated for the macroscopic measuring device $B$. Yet, contrary
to what may be suggested by Leggett and Garg's statements, for theories
where wMR (or wLR) is valid, we have argued that there is \emph{no
conflict }with the arguments of quantum measurement theory. This
is because system $B$ has a definite value $\lambda_{M}^{(B)}$ for
the outcome of $\hat{S}^{(B)}$. The system $A$ of $|\psi_{M}\rangle$
also has a definite value $\lambda_{M}^{(A)}$ for the outcome of
$\hat{\sigma}_{z}$ when prepared in (\ref{eq:ent-m}).

In short, contrary to what might be supposed, the argument that the
systems (\ref{eq:ent-m}) can be considered to have a definite real
property $\lambda_{M}^{(B)}$ and $\lambda_{M}^{(A)}$ does not contradict
the Bell violations (for example, of (\ref{eq:bell-chsh-1})), since
we have shown consistency with wMR (and also with wLR) for such violations.
The values however refer \emph{only} to systems prepared appropriately
at a given time  in a superposition with respect to pointer-bases
of $\hat{\sigma}_{z}^{(A)}$ and $\hat{S}^{(B)}$. Suppose one
could specify that $A$ (prior to the measurement interaction) were
prepared appropriately in $|\psi\rangle_{A}$ of (\ref{eq:sup}),
for the pointer-basis of $\hat{\sigma}_{z}^{(A)}$: According to
wLR, the system $A$ has a definite value $\lambda^{(A)}$ for the
outcome of $\hat{\sigma}_{z}^{(A)}$. Provided $\lambda^{(A)}=\lambda_{M}$,
it can then be argued that the system $|\psi_{M}\rangle$ is a suitable
measuring device.

There \emph{is} however a ``conflict'' as referred to by Leggett
and Garg, who refer to a ``\emph{definite macroscopic state}''.
The conflict arises when we consider the consequence of the wMR and
wLR assumptions, summarised in Section II, concerning the completeness
of quantum mechanics: The systems cannot be considered to be in either
quantum state $|\uparrow\rangle_{A}$ or $|\downarrow\rangle_{A}$,
or in $|\beta\rangle$ or $|-\beta\rangle$. If wMR is valid, it is
unclear what '\emph{state'} each of the systems are in? An analysis
of ontological states defining macroscopic realism has been given
by Maroney and Timpson \citep{maroney-timpson,Maroney}, which motivates
the following section.

\section{Comparison with other models of macroscopic realism}

Our conclusions are consistent with those of Maroney \citep{Maroney}
and Maroney and Timpson \citep{maroney-timpson}, who in analysing
tests of macrorealism have argued that violations of the Leggett-Garg
inequalities arise from a nonclassical form of measurement disturbance
and do not necessarily imply failure of macroscopic realism. Maroney
and Timpson considered three models of macroscopic realism, which
they refer to as macrorealism models. First, they defined \emph{operational
eigenstates of a property }as ``those preparations {[}of the system{]}
which determinately fix the value of the property''. In our context,
these are preparations of the system for which there is a predetermined
value for the outcome of the measurement $\hat{S}_{\theta}$.

The three models of macroscopic realism considered are: operational
eigenstate mixture macrorealism (OEM-MR); operational eigenstate support
macrorealism (OES-MR); and supra eigenstate support macrorealism (SES-MR).
Maroney and Timpson argued that only OEM-MR gives the strict form
of macrorealism that necessarily leads to the derivation of the Leggett-Garg
inequality. We next examine each of these models for consistency with
weak macroscopic realism as defined in this paper.

\subsection{Operational eigenstate mixture macrorealism}

The OEM-MR specifies that the system after preparation (after the
unitary interactions $U_{\theta}$) is in a mixture of the operational
eigenstates. This model is negated by the violation of Leggett-Garg
inequalities, and is compatible with wMR, but is a stronger model
than required by wMR.

An example of an OEM-MR model is the mixed state (\ref{eq:mix}),
where the system $A$ prior to the measurement $\hat{S}$ can be considered
to be with some probability either in the state $|\alpha\rangle$
or $|-\alpha\rangle$. The coherent states become quantum eigenstates
of $\hat{S}$ (for large $\alpha$), and are also operational eigenstates.
Here, $\hat{S}$ distinguishes between the two coherent states (for
large $\alpha$). The measurement can be shown to not change the system
placed in one or other coherent states.

Maroney analyses the three-box paradox, where MR would imply that
a ball placed in a superposition of being in one of three boxes is
always actually in one or other box \citep{Maroney}. A Condition
(III) is satisfied that a measurement made on the system where a ball
is placed in a box is confirmed to be non-disturbing to the state
of the system. This confirms that the measurement is non-invasive
for operational eigenstates. Maroney claims that ``\emph{An intuition
lurking alongside the idea that the ball is always in one, and only
in one, of the boxes, is that whenever the ball is in a given box,
it behaves exactly as it appears to behave when it is observed to
be in that box. This runs into difficulties, for when the ball's location
is observed, it is in an operational eigenstate. This rather natural
idea of macrorealism would lead to operational eigenstate mixture
macrorealism} ...''

Weak macroscopic realism (wMR) does not imply OEM-MR, since it is
not assumed that the state of the system before and after the measurement
are the same. This is evident from the analysis of Section II, where
it is proved for the cat state (\ref{eq:cat-1}) that, if wMR holds,
the system prior to the measurement $\hat{S}$ cannot be in one or
other quantum state that is an eigenstate of $\hat{S}$. The states
of the cat-system satisfying wMR are necessarily different before
and after the measurement.

\subsection{Operational eigenstate support macrorealism}

The OES-MR and SES-MR models consider the system to be, prior to measurement
$\hat{S}$, in a mixture of ontic states which have definite predetermined
values for the measurement $\hat{S}$. For the three-box paradox,
the measurement $\hat{S}$ corresponds to observing whether the ball
is found in a given box. Operational eigenstate support macrorealism
(OES-MR) constrains the ontic states to be in the support of the operational
eigenstates.

For OES-MR, Maroney comments about the application to the three-box
paradox \citep{Maroney}: ``\emph{The unobserved ball's ontic state
is always one that can occur when the ball is being observed. However,
the price is that those ontic states must now be behaving differently
to their appearances. Neither positive- nor negative-result non-invasiveness
will be possible, even for operational eigenstates. While the observed
behaviour of the ball, determinately placed in one box while Bob checks
Condition (III), is showing no detectable disturbance, something must
nevertheless be undergoing change, below the level of appearances,
as a result of Bob's measurements. This change takes place even when
Bob is only interacting with a different box: placing the ball in
Box 1, then opening the empty Box 2, somehow disturbs the ball in
Box 1 in an unobservable way. But when the system is prepared as in
a quantum superposition, and the ball is not being directly observed,
these same disturbances emerge and lead to observable consequences.}''

Positive-result and negative-result non-invasiveness refers to the
measurement having no disturbance to the system when the system is
directly measured as a ball being observed in a Box, and indirectly
measured, as in a ball not being observed in a Box. The OES-MR model
allows for nonlocality, since there can be a disturbance to the 'state'
of the ball in Box 1, when an empty Box 2 is observed.

Our work expands the analysis of Maroney for the OES-MR model. Here,
wMR posits that the observation of a ball not being in Box 2 would
not change the variable $\lambda_{M}^{(1)}$ that predetermines the
outcome of the measurement on the Box 1. However, the state of the
system can change. If there is a further unitary interaction at Box
1, and also at Box 2, so that measurement settings change, observable
paradoxes can occur.

We note that wMR counters OES-MR, since it is not true that ``the
unobserved ball's ontic state is always one that can occur when the
ball is being observed''. The ``observed'' state of the system
is identifiable as a quantum state, and for the cat state (\ref{eq:cat-1}),
we have seen that the assumption of wMR implies the system cannot
be in a quantum state prior to measurement.

\subsection{Supra eigenstate support macrorealism}

The supra eigenstate support macrorealism (SES-MR) model also considers
the system to be, prior to measurement $\hat{S}$, in a mixture of
ontic states which have definite predetermined values for the measurement
$\hat{S}$. Different to the OES-MR model however, the SES-MR model
allows novel ontic states that cannot be prepared quantum mechanically.

Maroney in examining the third SES-MR model states that: ``\emph{Supra
eigenstate support macrorealism takes the opposite route. Operational
eigenstates do not appear to be disturbed by Bob's measurements, and
it may be maintained that the ontic states in their support are not,
in fact disturbed. However, when the ball is prepared through a quantum
superposition, it may now be in an ontic state that does not appear
in any operational eigenstate. When it is not being observed, the
ball can behave differently.}''

The premise of wMR gives support to the SES-MR model of macroscopic
realism proposed by Maroney and Timpson. These authors also present
the de Broglie-Bohm model \citep{Bohm-1} as an example of an SES-MR
model \citep{Maroney}. In a recent paper \citep{q-weak}, the wMR
premises have been shown consistent with a model for realism based
on the $Q$ function \citep{q-contextual,simon-q,objective-fields-entropy}.
Analysis of that model suggests ontic states that cannot be compatible
with ``prepared'' or ``observed'' states \citep{q-measurement}.

\section{Discussion and conclusions}

In this paper, we have examined a macroscopic version of a Leggett-Garg
and Bell test presented earlier \citep{manushan-bell-cat-lg}, in
which the spin states $|\uparrow\rangle$ and $|\downarrow\rangle$
are realised by coherent states $|\alpha\rangle$ or $|-\alpha\rangle$,
with $\alpha\rightarrow\infty$, and the unitary interactions determining
the measurement settings $\theta$ in the Bell test, normally realised
by polarising beam splitters or Stern-Gerlach apparatuses, are realised
by local nonlinear interactions $U_{\theta}=e^{-iH_{NL}t/\hbar}$.
In particular, the set-up allows the noninvasive measurability premise
of the Leggett-Garg inequalities to be replaced by that of Bell's
locality assumption. The corresponding Bell test is macroscopic, meaning
that the Bell premises combine the assumptions of macroscopic realism
(MR) and locality at a macroscopic level (ML).

Earlier work showed how MR if defined \emph{deterministically} can
be falsified \citep{manushan-bell-cat-lg,jeong-macro-coarse-thermal-cat}.
Macroscopic realism applies to a system with two or more macroscopically
distinct states available to it, and assumes the system is in one
of those states, to the extent that a measurement $\hat{S}_{\theta}$
distinguishing between the states has a predetermined outcome. \emph{Deterministic
macroscopic realism} posits a predetermination of the outcome \emph{prior}
to the entire measurement dynamics, including the implementation of
$U_{\theta}$, and is a stronger (more restrictive) assumption. In
such a model, as in classical mechanics, it is assumed there are a
set of macroscopically distinct states giving a definite outcome for
$\hat{S}_{\theta}$, which can be identified for the system prior
to the time at which $U_{\theta}$ is implemented.

Violations of Bell inequalities are explained generally as a failure
of ``local realism'', or of ``local hidden variables'' \citep{Bell-2}.
The violations exclude that there can be hidden variables satisfying
the Einstein-Podolsky-Rosen (EPR) premises. EPR's ``elements of reality''
are negated by Bell violations. The macroscopic version of the Bell
test motivates a deeper consideration of the meaning of local realism
and, in particular, of the EPR premises, since any rejection of macroscopic
realism would be a more startling conclusion than the rejection of
local realism at the microscopic level.

Our conclusion is that MR is not contradicted by the Bell violations,
and can be viewed consistently with the violations, if defined in
a less restrictive way, as \emph{weak macroscopic realism} (wMR).
Weak macroscopic realism has been proposed earlier, and recent work
gives an extension of the definition to the bipartite set-up of EPR
\citep{ghz-cat}. The earlier work showed consistency of the macroscopic
Bell violations with a subset of the wMR premises \citep{manushan-bell-cat-lg}.
Here, we show consistency of the macroscopic Bell violations with
the full definition of wMR. Ref. \citep{manushan-bell-cat-lg} proposed
three tests of wMR, where the results would be consistent with wMR
according to quantum mechanics. We extend to present a fourth test,
involving EPR's ``elements of reality''.

The consequence of our work is a model consistent with quantum mechanics,
in which there is an understanding of when the EPR ``elements of
reality'' can be considered to apply. The ``elements of reality''
will apply to the system defined after the unitary interaction $U_{\theta}$
has been carried out in the experiment. We see that the violation
of the Bell inequalities occurs due to a \emph{combination} of a failure
of realism and locality. On the other hand, both a weaker version
of realism and a weaker version of locality apply: The system has
a real property for the outcome of the measurement $\hat{S}_{\theta}$
after the implementation of $U_{\theta}$. Also, the system has an
``element of reality'' for the outcome of $\hat{S}_{\theta}$, \emph{if
}the outcome of $\hat{S}_{\theta}$ at $A$ can be predicted with
certainty by a measurement $\hat{S}_{\phi}$ on a second system $B$
$-$ but this applies only \emph{once} the implementation of the unitary
interaction $U_{\phi}$ at $B$ has taken place.

A justification for wMR is given on considering the nature of quantum
measurement. Consider a system $A$ for which $\hat{S}_{\theta}$
is being measured, by a coupling to a macroscopic meter, system $B$.
This is a situation for which EPR's ``element of reality'' apply,
because one can predict with certainty the outcome of the measurement
on system $A$ by performing a measurement on the meter $B$. While
EPR's traditional ``elements of reality'' can be negated, this particular
``element of reality'' is justified by wMR, because the coupling
interaction is such that the measurement basis $\theta$ has already
been specified. Hence, wMR resolves paradoxes about macroscopic realism
and measurement, as highlighted by Leggett and Garg \citep{legggarg-1}.

While the motivation for proposing that wMR is valid is to arrive
at a model allowing some form of macroscopic reality, the mapping
between the microscopic and macroscopic Bell tests ensures that a
similar definition, \emph{weak local realism} (wLR) \citep{ghz-cat},
can be applied to the original set-up involving spin states $|\uparrow\rangle$
and $|\downarrow\rangle$. The original Bell violations can be explained
consistently with wLR. A justification for wLR can also be given based
on the argument that in the microscopic tests at the time after the
unitary dynamics $U_{\theta}$ establishing the measurement setting
$\theta$, there will be some form of amplification, such as a coupling
to a meter \citep{ghz-cat}. Hence, wMR can be applied at this time.

It is interesting to consider the possibility of an experiment. While
the predictions of wMR are consistent with those of quantum mechanics,
four tests of wMR have been presented, which motivates an experiment.
Two-mode entangled cat states have been experimentally realised \citep{cat-bell-wang}.
However, it is challenging to realise $U_{\theta}$. The Bell example
with cat states was presented because of the strength of the conclusions
that follow from a Bell violation, and because of the simplicity of
the argument from a theoretical viewpoint. Other macroscopic realisations
of quantum correlations can be considered however \citep{macro-review}.
This includes the continuous variable correlations of the Einstein-Podolsky-Rosen
(EPR) paradox which are measured by homodyne detection, the measurement
setting $\theta$ being a phase shift \citep{epr-r2,epr-rmp}. Here,
set-ups are possible where amplification takes place prior to the
implementation of the phase shift $\theta$ \citep{quant-qi-mdr,cv-bell-macro,bowen-epr},
so that macroscopic states can be defined and both dMR and wMR posited
for the system. A study of wMR for such an experiment would lead to
the possibility of tests of wMR, along the lines proposed for Bohm's
version of the EPR paradox in \citep{ghz-cat}. We also note that
mesoscopic quantum correlations have been achieved for atomic systems
\citep{treutlein-exp-bell-1,tura,eprnaturecommun-2-1,bell-kasevich-1,SteerAt-obert-1-1,treu-matteo-1-1,EntAtoms-1-1,bryan-reviews-1,bryanlibby-1,andrei-steer}.
In particular, EPR correlations involving atomic clouds have been
measured, including where the measurement setting is adjustable locally
\citep{sch-epr-exp-atom}. This has led to a realisation of Schrödinger's
description of the EPR paradox, in which there is a simultaneous measurement
of two non-commuting observables, $x$ and $p$ \citep{s-cat-1935}.
An analysis of wMR showing consistency for such EPR correlations would
seem possible.

In conclusion, we have outlined how a weak form of local realism can
be consistent with realism at a macroscopic level, despite violations
of macroscopic Bell inequalities. Yet, the argument presented by Schrödinger
is that there is inconsistency between (weak) macroscopic realism
and the completeness of quantum mechanics \citep{s-cat-1}: for a
macroscopic superposition, if there is macroscopic realism, then what
state is the system in prior to detection $-$ the system cannot be
viewed as being in any quantum state? This motivates analysis of deeper
models or interpretations of quantum mechanics \citep{bohm-hv,philippe-grang-context,hall-2}.
As one example, a model for field amplitudes based on the Q function
shows how consistency between quantum mechanics and weak macroscopic
realism might be achieved using microscopic retrocausal fields \citep{q-contextual,objective-fields-entropy,simon-q}.
\begin{acknowledgments}
This research has been supported by the Australian Research Council
Discovery Project Grants schemes under Grant DP180102470 and DP190101480.
We thank NTT Research for technical help and motivation for this project.
We are grateful for support from the Templeton Foundation.\textcolor{blue}{}
\end{acknowledgments}

\end{document}